\documentclass[%
 reprint,
 amsmath,amssymb,
prb,
]{revtex4-2}
\usepackage[english]{babel}
\usepackage{subcaption}
\usepackage{graphicx}% Include figure files
\usepackage{comment}
\usepackage{dcolumn}% Align table columns on decimal point
\usepackage{bm}% bold math
\usepackage{hyperref}% add hypertext capabilities
\hypersetup
{
colorlinks=true,
linkcolor=blue,
citecolor=blue,
filecolor=blue,
urlcolor=blue
}
\usepackage{dsfont}
\usepackage[noabbrev]{cleveref}
\usepackage{caption}
\usepackage{subcaption}
\usepackage{overpic}

\usepackage[font=small,justification=justified,singlelinecheck=false]{caption}
\usepackage{ragged2e}
\begin{document}

\preprint{APS/123-QED}

\title{Topological phases and Edge states in an exactly solvable\\ Gamma matrix model}% Force line breaks with \\

\author{Akhil Pravin Furtado}
\email{akhil.furtado@gmail.com}
\author{Kusum Dhochak}%
 \email{kdhochak.iitpkd.ac.in}
\affiliation{%
Department of Physics, Indian Institute of Technology Palakkad, Palakkad, Kerala, 678623, India\\
}%


%\affiliation{
%\affiliation{
 %Third institution, the second for Charlie %Author
%\affiliation{%



\begin{abstract}
%This paper extends a previous work (see Ref.~\cite{PhysRevE.106.024114}) on a 1-D quantum many-body model with $\Gamma$ matrix ($2^d\times2^d$ dim) degrees of freedom. 
%where the system Hamiltonian, energy band structure and a few symmetries were discussed. 
%While the earlier study provided a general understanding of the system, this study aims to explore and analyze the non-trivial topological phases, topological phase transitions, and other subtleties. 

We study the phases of an exactly solvable one dimensional model with $4-$dimensional $\Gamma-$matrix degrees of freedom on each site. The $\Gamma-$matrix model has a large set of competing interactions and displays a rich phase diagram with critical lines and multi-critical points. We work with the model with certain $Z_2$ symmetries and identify the allowed symmetry protected topological phases using the winding number as the topological invariant. The model belongs to the CII-class of the $10-$fold classification and allows for integer values of the winding number. We confirm that the system also hosts localized zero energy Majorana edge modes, consistent with the integer value of the winding number of the corresponding phase. We further study scaling and universality behaviour of the various topological phase transitions.
\\
\end{abstract}

\maketitle

%\tableofcontents
\section{\label{sec:Introduction}Introduction\protect}
\hspace*{2mm}

%%%%%%%%%%%%%%%%%%%%%%%%%%%%%%%%%%%%%%%%%%%%
Exactly solvable models are extensively used to understand diverse phenomena and quantum critical behaviour in many-body systems (Refs.\cite{Altland2023-sa,Sachdev2011-gv,Dutta2015-ke,PhysRevLett.25.443,PFEUTY197079,Franchini_2017,10.1119/1.1970340}).  These also often serve as test beds for newer theoretical and numerical techniques. In recent decades, a lot of focus has been on exploring topological phases of matter (Refs.\cite{XGWEN,RevModPhys.88.035005,Asb_th_2016,PhysRevB.86.205119}), their non trivial particle excitations (Refs.\cite{Kitaev_2009,PhysRevB.83.035107,PhysRevB.85.035110,Ryu_2010,Chen_2013,Senthil_2013,PhysRevB.78.195125}) and their relevance for quantum computation and quantum information (Refs.\cite{RevModPhys.80.1083,KITAEV20032,Sarma_2015,PhysRevLett.97.180501,PhysRevB.95.235305,PhysRevB.90.155447,PhysRevLett.100.096407}).  One dimensional fermion chains and spin chains have been a considerable part of these studies owing to their relative analytical and numerical tractability. These models have also been extensively studied as candidates for their topological phases as well as for creating and manipulating topological excitations like Majorana Fermions, magnons, parafermions, anyons etc (Refs.\cite{AYuKitaev_2001,Alicea_2012,Clarke_2013,Fendley_2012,PhysRevB.81.134509,PhysRevB.83.075102,PhysRevB.87.174427}). 
These one-dimensional models are used to characterize different types of systems like proximity induced superconducting chains, spin chains, cold atom systems, as well as quasi one-dimensional materials (Refs.\cite{PhysRevLett.105.077001,doi:10.1126/science.1222360,PhysRevX.4.031027,PhysRevLett.105.177002,Atala_2013,Meier_2016}).

\begin{figure}[!h]
\includegraphics[scale=0.4]{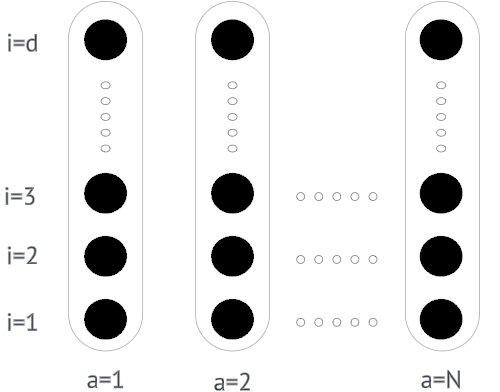}
       
        \caption{\justifying Illustration of the Lattice/Sub-lattice structure for the model with the lattice sites (vertical blocks) labeled by the index $a$, and the sublattices (solid black circles) labeled by the index $i$. %Here, $a\equiv\text{lattice}\,\,,\text{and}\,\,i\equiv \text{sublattice/qubit}$ indice 
        }
            \label{fig:Latticerep1}
        \end{figure}

Several one-dimensional models, such as the Kitaev chain (Ref.\cite{AYuKitaev_2001}), coupled Kitaev chains (Refs.\cite{PhysRevB.89.174514,WU20123530,ZHOU20172426}), and long-range spin models (Ref.\cite{PhysRevB.85.035110}), are known to host topologically non-trivial phases with topologically protected Majorana zero energy modes at the boundaries. Various works have explored the ways to increase the number of distinct topological phases and protected edge modes in one-dimensional systems using coupled chains, extended-range interactions and higher spin configurations (Refs.\cite{PhysRevE.106.024114,PhysRevB.108.094411,PhysRevD.110.094510}). The $\Gamma-$matrix degrees of freedom in this study naturally provide a higher-dimensional Hilbert space per site, leading to possibilities of a richer phase diagram and a larger number of Majorana modes in a single chain with nearest neighbour interactions. The enlarged on-site Hilbert space can be interpreted in various ways—either as representing higher spin states or as comprising multiple pseudospins or sublattices per site (Fig.(\ref{fig:Latticerep1})), and possibly capturing properties of phase diagrams of spin ladders while having the advantage of exact solvability.
The $\Gamma$ matrix model studied in Ref. \cite{PhysRevE.106.024114} is exactly solvable for $2^d$ dimensional $\Gamma$ matrices. In this work, we focus on the simplest non-trivial case of $ 4 \times 4$  matrices and explore the topological phases of this model, critical behaviour, and the nature of the zero-energy symmetry-protected edge modes. 
%among themselves.
 \newline
\hspace*{2mm}The rest of the paper paper is organized as follows: in Sec.~\ref{sec: The Model},  we review the model and the exact solution, discuss the symmetries of the system and define the topological invariant. 
In Sec.~\ref{sec:Phase Diagrams}, we study the phase diagrams, topological classification of the phases and the zero energy topologically protected edge modes. 
We then analyze the nature of the critical points and the universality classes via the scaling behavior in sec.~\ref{sec:Critical Behaviour}. 

\section{\label{sec: The Model}Symmetries and topological classification}
%%%%%%%%%%%%%%%%%%%%%%%%%%%%%%%%%%%%%%%%%%%%%%%%%%%%%%%%%%%%%%%%
\subsection{\label{subsec:systems H} Review of System Hamiltonian}
\hspace*{2mm }
The general model \cite{PhysRevE.106.024114} is defined as a one dimensional lattice model with $\Gamma-$ matrix operators on each site with nearest neighbour interactions and a local external field (Eq.(\ref{Gammarealhamiltonian})). Each site hosts $2d+1$ mutually anticommuting $\Gamma-$ matrices ($d=1,2..$). These $\Gamma$ matrices generate a Euclidean Clifford algebra $Cl_{2d+1}(\mathbb{R})$  \cite{Rudolph2017,PhysRevB.108.094411}.
\begin{equation}
      H=-i\sum_{a=1}^{N}\sum_{\mu,\nu=1}^{2d}J_{\mu\nu}\Gamma^{\mu}_a\Gamma^{2d+1}_a 
\Gamma^{\nu}_{a+1}-\sum_{a=1}^{N}\sum_{i=1}^{d}h_iS^i_a\label{Gammarealhamiltonian}
\end{equation}
Here $a$ labels the lattice sites and $\mu,\nu\in[1,2d]$ are the $\Gamma-$matrix indices \& $J_{\mu\nu}$'s are the nearest neighbour interaction parameters. Also, $\Gamma^{2d+1}=(-i)^d\prod_{\mu=1}^{2d}\Gamma^{\mu}$ \textit{s.t.}
\begin{equation}
    \{\Gamma^{\mu}_a,\Gamma^{\nu}_a\}=2\delta^{\mu,\nu}\,;\,\{\Gamma_a^{\mu},\Gamma_a^{2d+1}\}=0\,;\,(\Gamma_a^{2d+1})^2=\mathds{I}%_{2^d\times2^d}.
\end{equation}
 The local `spin' operator $S_i$  is defined as, $S_i=-i\Gamma^{2i-1}\Gamma^{2i}$ with $i=1,2..,d$ and $h_i$ is an external field.\\
\hspace*{2mm}For $d=1$, these reduce to the Pauli $\sigma$ matrices.
The model is exactly solvable for all $d$, but in this work we focus on the phases of the $d=2$ system with $ 4$-dimensional matrices on each lattice site.

These $4-$dimensional $\Gamma-$matrices can be represented with a direct product of two spin-$1/2$ operators $\vec{\sigma_1}, \vec{\sigma_2} $. The representation is not unique, and we use the following form for the $\Gamma$-matrices,
%\begin{center}
%$\Gamma^{1}=\sigma^1_1\otimes\mathds{I}_2\otimes\mathds{I}_3...\otimes\mathds{I}_d $\\
    %$\Gamma^{2}=\sigma^2_1\otimes\mathds{I}_2\otimes\mathds{I}_3...\otimes\mathds{I}_d$\\%
%$\Gamma^{3}=\sigma^3_1\otimes\sigma^1_2\otimes\mathds{I}_3...\otimes\mathds{I}_d $\\
%$\vdots$

%$\Gamma^{2d-1}=\sigma^3_1\otimes\sigma^3_2\otimes\sigma^3_3...\otimes\sigma^1_d $\\
%$\Gamma^{2d}=\sigma^3_1\otimes\sigma^3_2\otimes\sigma^3_3...\otimes\sigma^2_d
%$\newline\newline
%$\Gamma^{2d+1}=(-i)^d\prod_{\mu=1}^{2d}\Gamma^{\mu}=\sigma^3_1\otimes\sigma^3_2\otimes\sigma^3_3...\otimes\sigma^3_d
%$
%\end{center}
%This is the choice of the general representation adopted in this work. But in the later sections, we work with d=2.
\begin{center}
$\Gamma^{1}=\sigma^1_1\otimes\mathds{I}_2\,;$\\ $\Gamma^{2}=\sigma^2_1\otimes\mathds{I}_2\,;$\\
$\Gamma^{3}=\sigma^3_1\otimes\sigma^1_2\,;$\\ $\Gamma^{4}=\sigma^3_1\otimes\sigma^2_2\,;$
\vspace{-0.3cm}
\begin{equation}
\Gamma^{5}=\sigma^3_1\otimes\sigma^3_2;\,\, S_1=\sigma_1^3\otimes\mathds{I}_2\,;
    S_2 = \mathds{I}_1\otimes\sigma_2^3.
\label{repsigma}  
\end{equation}
\end{center}
%\subsection{Energy Spectrum}
The model can be  exactly diagonalized by the Jordan-Wigner (JW) fermionization mapping the $\Gamma-$matrices to Majorana Fermions ($\chi^{\mu}$) \cite{PhysRevE.106.024114}.
\begin{equation}
    \chi^\mu_a=(\prod_{b=1}^{a-1}\Gamma^{5}_b)\,\Gamma_{a}^{\mu}\,\,\,;\,\,\,\mu\in[1,4]
\end{equation}
with $\,\,\chi^{\mu}_a=(\chi^{\mu}_a)^{\dag}\,\,\,\,\text{and }\,\,\,\, \{\chi^{\mu}_a,\chi^{\nu}_b\}=2\delta^{\mu\nu}\delta_{ab}$. In the thermodynamic limit with  periodic boundary conditions (PBC) (Ref.{\cite{PhysRevE.106.024114}}), the momentum-space representation can be defined as  $\chi^{\mu}$'s is,\begin{equation}
     \chi^{\mu}_{a}=\frac{1}{\sqrt{N}}\sum_{k\in 1^{st}\text{BZ}} \,\,e^{ika}\chi^{\mu}_{k}
\,\end{equation} with $(\chi^{\mu}_{k})^{\dag}= \chi^{\mu}_{-k}\,\,\,
\,\&\,\,\,\{ \chi^{\mu}_{k}, \chi^{\nu}_{k'}\}=2\delta^{\mu\nu}\delta_{kk'}$ and $k,k'\in$ $1^{st}$ Brillouin zone (BZ). Writing $J_{\mu\nu}=\frac{1}{2}(S_{\mu\nu}-A_{\mu\nu}),$ the quadratic Hamiltonian in momentum-space is,
\begin{multline}
      H=\sum_{k>0}(\sum_{\mu,\nu=1}^{4}(i\,A_{\mu\nu}\,cos(k)+S_{\mu\nu}\,sin(k)
    )\chi^{\mu}_{-k}\chi^{\nu}_{k}\\+i\sum_{i=1}^{2} h_i(\chi^{2i-1}_{-k}\chi^{2i}_{k}-\chi^{2i}_{-k}\chi^{2i-1}_{k})\label{Majorana k>0 hamiltonian}
\end{multline}
\hspace*{2mm}%As stated earlier, we set d=2 and 
The model has 16 independent parameters to work with.
%In the Sec.(\ref{subsec:Discrete symmtries}), we impose certain discrete symmetries and restrict all future discussions to a smaller parameter sub-space.

%%%%%%%%%%%%%%%%%%%%%%%%%%%%%%%%%%%%
\subsection{\label{subsec:Discrete symmtries}Discrete Symmetries}
Symmetries of the system, if present, are useful  in defining topological invariants and and classifying the phase diagram. Since we have a large number of parameters to work with, we also use this freedom to work in the parameter regimes where the system has various important  symmetries. The particular symmetries that we analyse are as below:
\paragraph{
 General Symmetry:} The only general discrete symmetry that the system possesses is:
\begin{equation}
    \Gamma^{\mu} \longrightarrow -\Gamma^{\mu}\label{inversiongamma}
\end{equation}
%This symmetry leads to the invariance of the topological order of the parameter spaces when $h\to-h$ (see Sec(\ref{sec:Phase Diagrams})).
\paragraph{Spin-reflection symmetry:} This symmetry is the exchange of $\Gamma$'s at $a^{th}$ site with that of the $(N+1-a)^{th}$ site.
\begin{equation}
    \Gamma_a^{2i-1}=-i\,\Gamma^{2i}_{N+1-a}\Gamma^{2d+1}_{N+1-a}\,\,;\,\,\Gamma_a^{2i}=i\,\Gamma^{2i-1}_{N+1-a}\Gamma^{2d+1}_{N+1-a}
\end{equation}
and,
\begin{equation}
    S^i_a\longrightarrow S^i_{N+1-a}\,\,;\,\,\Gamma^{2d+1}_a\longrightarrow \Gamma^{2d+1}_{N+1-a}
\end{equation}
For this to be a symmetry for the model, we impose the following constraints on the parameters,
\begin{equation}
\begin{array}{c c}
     S_{2i,2j}=-S_{2i-1,2j-1}&S_{2i,2j-1}=S_{2i-1,2j}\\
     A_{2i,2j}=A_{2i-1,2j-1}&A_{2i,2j-1}=-A_{2i-1,2j}
\end{array}\label{reflection}
\end{equation}\newline
\paragraph{ Particle-hole Symmetry ($\hat{P}$):} It is an anti-unitary transformation and anti-commutes with the Hamiltonian. In the majorana fermion basis (Eqn.(\ref{majbasis})),%The Complex Fermion (CF) basis (see Appendix\ref{app:Complexfermionlanguage}) provides a mainstream understanding and in this basis,P= -i\kappa(\sigma_y\otimes\mathds{I})
\begin{equation}
   P=\kappa U_p=\kappa(\mathds{I}\otimes\sigma_y) \,\,;\,\,P^2=-\mathds{I}\label{particlehole}
\end{equation}
Here, $\kappa$ is the complex conjugation operator (Ref.\cite{PhysRevE.106.024114}). 
\paragraph{ Chiral Symmetry ($\hat{S}$):} It is a unitary transformation and if we impose the constraints $S_{14}=0=A_{13}$, it anti-commutes with the Hamiltonian. It plays an important role in the topological classification of the phases (Sec.(\ref{subsec:Winding Number})) and we work with this constraint in the rest of the paper. In the Majorana fermion basis,
\begin{equation}
    S=(\mathds{I}\otimes\sigma_x)\,\,;\,\,S^{2}=\mathds{I}\label{chiral}
\end{equation}
\paragraph{ Time reversal symmetry ($\hat{T}$):} It is an anti-unitary transformation and commutes with the Hamiltonian. Since both $\hat{P}$ and $\hat{S}$  have been imposed $\hat{T}$ exists and $T= S P^{-1}$. In the Majorana Fermion basis,
\begin{equation}
    T=i(\mathds{I}\otimes\sigma_z)\kappa\,\,;\,\,T^{2}=-\mathds{I}\label{timerev}
\end{equation}
\newline
With these symmetries, the system belongs to the CII symmetry class of the Ten-Fold classification (Ref.\cite{PhysRevB.55.1142, Ryu_2010}), allowing the system's topological invariant (TI) to take integer values, $TI\in\mathds{Z}$.

%%%%%%%%%%%%%%%%%%%%%%%%%%%%%%%%%%%%
\subsection{\label{subsec:Gap closing}Diagonalization and Phase Transitions}
%\textcolor{red}{d=2 matrix, gap, energy bands, gap closing conditions for k=0 state, how does it relate to $k=\pi$,
%then write it for k incommensurate, state about finding any %critical parameter.}\newline
Working with  the above symmetries, the Hamiltonian now has the form: 
\begin{equation}
    H=\sum_{k>0}V^{\dag} h(k) V \,\,;\,\,V=(\chi^1_k,\chi^2_k,\chi^3_k,\chi^4_k)^T\label{majbasis}
\end{equation}
with,
\begin{widetext}
\small
\begin{equation}
h(k)=\begin{pmatrix}
         S_{11}\,Sin(k) & i(h_1 + A_{12} \,Cos(k)) & S_{13} \,Sin(k) & 
 i\,A_{14} \,Cos(k) \\
 -i(h_1 + A_{12}\, Cos(k))& -S_{11}\, 
  Sin(k) & -i\, A_{14}\, Cos(k)& 
 - S_{13}\, Sin(k)\\  S_{13}\, Sin(k)& 
 i\, A_{14}\, Cos(k) & S_{33}\, Sin(k)& 
 i(h_2 + A_{34}\, Cos(k)) \\
 -i \,A_{14}\, Cos(k) & - 
  S_{13}\, Sin(k)& -i(h_2 + A_{34}\, Cos(k))& -S_{33}\, Sin(k)
    \end{pmatrix}.\label{h(k)}
\end{equation}
\end{widetext}
\normalsize
The spectrum has four energy bands (Ref.\cite{PhysRevE.106.024114}),  $\pm\epsilon_{+}(k),\pm\epsilon_{-}(k),$ with 
\begin{equation}
\epsilon_{\pm}(k)=\sqrt{F_k\pm\sqrt{F_k^2-G(k)}}.\,\,\, 
\end{equation}
Here, $G(k)=Det(h(k))$ and, \begin{multline*}
    F_k=\frac{1}{2}[(h_1+A_{12}\,\cos(k))^2+(h_2+A_{34}\,\cos(k))^2\\+2A_{14}^2\cos^2(k)+(S^2_{11}+S^2_{33}+2\,S_{13}^2)\sin^2(k)]
\end{multline*}
With both $F_k$ and $G(k)$ positive, the excitation energy gap is $2\epsilon_-$. The system can have $2$nd order phase transition when $\epsilon_- \rightarrow 0$ which happens when $G(k)=0.$ The ground state energy is given by 

\begin{equation}
    E_{g}=-\sum_{k>0}(\epsilon_{+}+\epsilon_{-})\label{GS1}
\end{equation}

We find possibilities of gap closing at commensurate momenta ($k=0,\pm\pi$) as well as for some incommensurate momenta as we vary differnet parameters of the model. 
%The energy band-gap is, \begin{equation}\Delta \epsilon=2\epsilon_{-}
%\end{equation}
%The system undergoes a quantum phase transition (QPT) when \begin{equation}\epsilon_{-}(A_{\mu\nu},S_{\mu\nu},h,k)=0\,\,;\,\, \text{ie. when}\,\, G_k= 0\end{equation}
%\hspace{2mm}Leading to two distinct cases of gap closing: 1) Commensurate k gap-closing ($k=0,\pm\pi$) and 2) Incommensurate k gap-closing.\newline
\hspace*{2mm}  Reparameterizing  $h_1=\mu\,h$ and $h_2=\mu^{-1} h$, The critical fields for the commensurate case of $k=0$ ( Ref.\cite{PhysRevE.106.024114}) are 
\begin{multline}
     h_{c1/2}=\frac{1}{2\mu}\,\,[\,\,(\mu ^2 A_{34}+A_{12})\pm\\\sqrt{\left(\mu ^2 A_{34}+A_{12}\right){}^2+4 \mu^2  \left(A_{14}^2-A_{34} A_{12}\right)}\,\,]\label{eqn:criticalh}
\end{multline}
and for the case of $k=\pm\pi$ we found the critical field values to be $-h_{c1/c2}$.\newline
\hspace*{2mm} 
For the  incommensurate $k$ transitions, the solutions are found self-consistently numerically. In presence of an anti-commuting  symmetry (eg. chiral symmetry  $\hat{S}$), the Hamiltonian can be transformed to an off-diagonal matrix by a unitary transformation $U_{od}$ constructed from the eigen vectors of the chiral symmetry \cite{Maiellaro_2019,PhysRevB.107.075422,ZHOU20172426}. 
\begin{equation}
    H_{od}=U_{od}^{\dag}h(k)U_{od}=\begin{pmatrix}
    0&A_{2\times2}(k)\\
    A^{\dag}_{2\times2}(k)&0
\end{pmatrix}\label{offdiagonal}
\end{equation}

This simplifies the gap closing condition as $G(k)=|\text{det}A|^2=0$ factorizes the quartic equation for the phase transition to two quadratic conditions $Re(\text{det} A)=0$ and $Im(\text{det} A)=0$.

%We can simplify the similar analytical calculation for the %criticality condition for 
%the incommensurate k case with a symmetry argument. If the system has a discrete unitary symmetry that anti-commutes with the Hamiltonian (eg: chiral symmetry), then one can block off-diagonalize the Hamiltonian using a unitary transformation ( $U_{OD}$). Here $U_{OD}$ is constructed from the eigenvectors of such a symmetry operation (Refs. \cite{Maiellaro_2019,PhysRevB.107.075422,ZHOU20172426}).
%Consequently, G(k)=$|det(A)|^2=0$ is the condition for the gap closing. This requires simultanouesly solving for two conditions Re(det(A))=0 and Im(det(A))=0 (both $\mathcal{O}(h^2)$) instead of solving a polynomial $\mathcal{O}(h^4)$ for $G_k=0$.
%%%%%%%%%%%%%%%%%%%%%%%%%%%%%%%%%%%%
\subsection{\label{subsec:Winding Number}Winding Number}
\hspace*{2mm}
%Since the system belongs to the CII class, 
The off diagonalization of the Hamiltonian in presence of the chiral symmetry also allows for defining a topological invariant, winding number ($W$), for the system using the off-diagonal block $A$ of the Hamiltonian (eqn.\ref{offdiagonal}). The winding number is defined as \cite{Maiellaro_2019,PhysRevB.107.075422,ZHOU20172426}
\begin{equation}
    W=\frac{1}{2\pi i}\oint_{1^{st} BZ} dk \, \frac{\partial\, ln(det(A(k))}{\partial k}. \,\,\,\,
\end{equation}
This is also the primary reason for imposing the additional constraints on coupling constants of the model (\textit{i.e.} $A_{14}=0=S_{13}$), to introduce the chiral symmetry discussed above. 

%Thus, the imposed $\mathds{Z}_2$ symmetries (Sec.\ref{subsec:Discrete symmtries}) play an important role in the topological classification of the phase (Sec.(\ref{sec:Phase Diagrams})).

%%%%%%%%%%%%%%%%%%%%%%%%%%%%%%%%%%%%%%%%%%%%%%%%%%%%%%%%%%%%%%%%
\section{\label{sec:Phase Diagrams}Phase Diagrams}
\begin{figure}
    \begin{subfigure}{0.375\textwidth}    \includegraphics[width=\textwidth,scale=0.3]{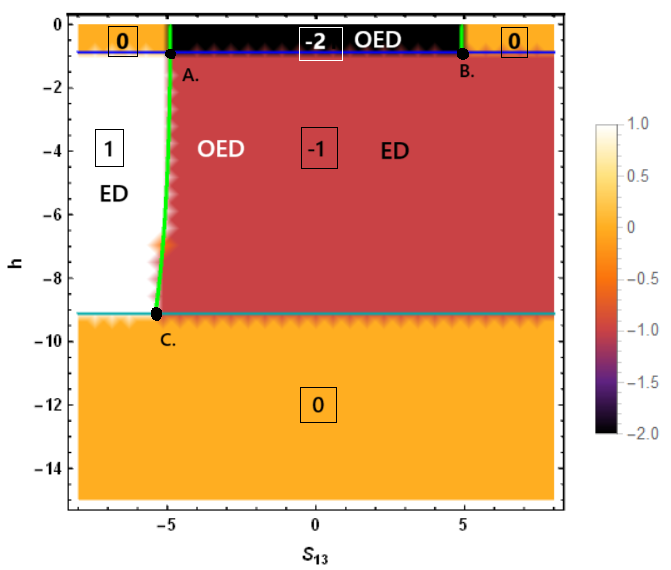}
    \caption{$A_{14}$= 4.}% Plot of W in the   Curves: Red: $h_{c1}$(k=0), Blue: $h_{c2}$ (k=0), Neon Green: incommensurate k. MCPs: A, B.
    \label{fig:a14=4}
\end{subfigure}
\begin{subfigure}{0.375\textwidth}
\includegraphics[width=\textwidth,scale=0.3]{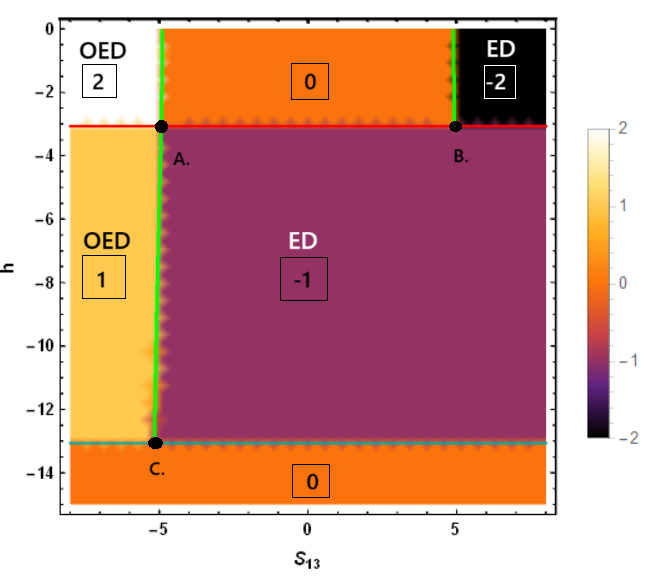}
    \caption{$A_{14}$= 8.} 
    %Curves: Red: -$h_{c1}$ (k=$\pm\pi$), Blue: $h_{c2}$(k=0), Neon Green: incommensurate k. MCPs: A, B,C}
    \label{fig:a14=8}
\end{subfigure}
%\captionsetup{width=0.5\textwidth} % Set the total caption width
    \caption{\justifying Phase diagram in the \ \( S_{13} \)-\( h \)  plane, with fixed $A_{14}=4, 8$ for (a) and (b) respectively. The colors and numbers inside the boxes represent the value of \( W \). Blue, Cyan (\( k = 0 \)), and Green (incommensurate \( k \)) are the critical lines. The muli-critical points are labeled as A, B, and C.  The text ED (exponential decay) and OED (oscillatory exponential decay) in the phases refers to the nature of localization of Majorana zero energy edge states (discussed in \ref{subsec:Spatial Distribution of MZE}).}
        \label{fig:a14merged}
    \end{figure}

\begin{figure}
    \begin{subfigure}{0.375\textwidth}
    \includegraphics[width=\textwidth,scale=0.3]{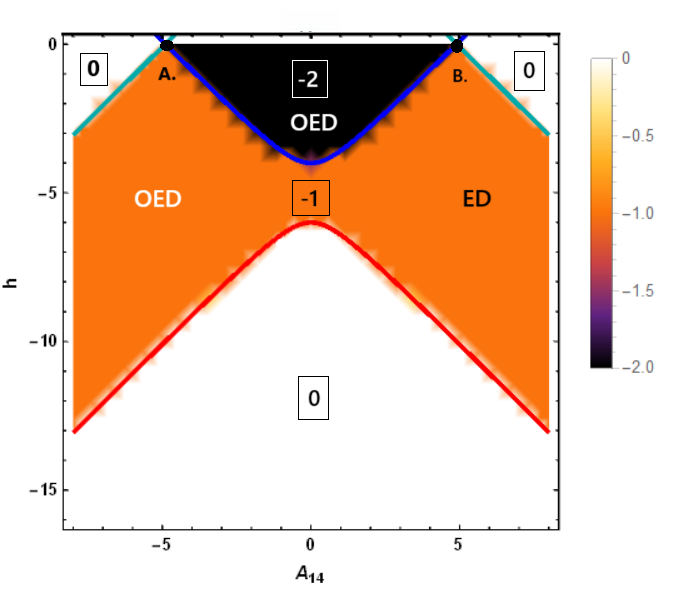}
    \caption{$S_{13}$= 4.}% Curves: Green: $-h_{c1}$(k=$\pm\pi$), Blue: $h_{c1}$ (k=0), Red: $h_{c2}$ (k=0). MCPs: A, B.
    \label{fig:s13=4}
\end{subfigure}
\begin{subfigure}{0.375\textwidth}
    \includegraphics[width=\textwidth,scale=0.3]{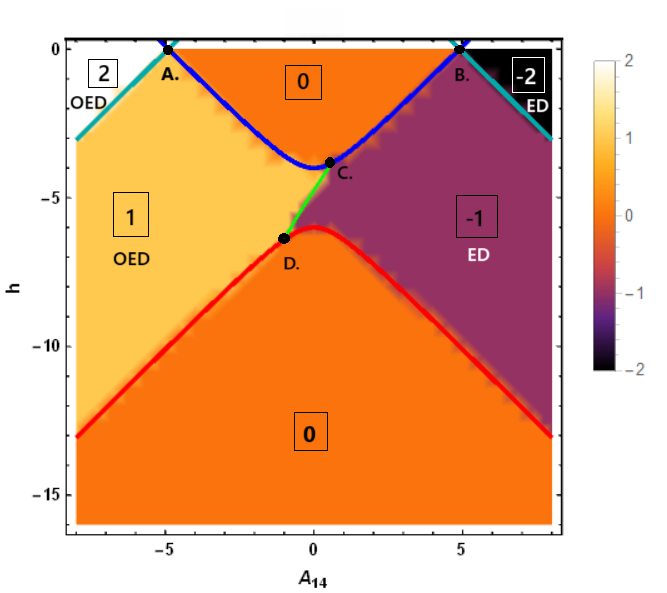}
    \caption{ $S_{13}$= 8.} %Curves: Red: -$h_{c1}$ (k=$\pm\pi$), Blue: $h_{c1}$(k=0), Green: $-h_{c2}$(k=0), Black: incommensurate k. MCPs: A, B, C and D}
    \label{fig:s13=8}
\end{subfigure}
      \caption{\justifying Phase diagram in the \( A_{14} \) -\ \( h \) plane, with fixed \(S_{13}=4,8\). The colors and numbers inside the boxes represent the value of \( W \). Blue, Red (\(k=0)\), Cyan (\(k=\pm\ \pi/a)\)), and Green (incommensurate \( k \)) are the critical curves. The Multi critical points are labeled with the letters A, B, C and D. The text ED (exponential decay) and OED (oscillatory exponential decay) in the phases refers to the nature of localization of Majorana zero energy edge states (discussed in \ref{subsec:Spatial Distribution of MZE}).}
        \label{fig:s13merged}
    \end{figure}

\hspace*{2mm} The Hamiltonian has a large number of parameters and we can expect several different phases as we vary the coupling parameters of the system. The system can also be thought of as a model of two coupled spin chains from the representation in eqn.(\ref{repsigma}) or as two interacting Kitaev chains (eqn. (\ref{h(k)})). Since individual Kitaev chains are well understood, to understand the new phases that can appear in the model, we fix the intra-chain coupling parameters and vary the inter-chain parameters $A_{14}, S_{13}$ along with the field $h$ and explore the phase diagram. We fix $A_{12}=4=S_{11}$ (chain-1 labeled as $C_1$), $A_{34}=6=S_{33}$ (chain-2  labeled as $C_2$) and $\mu=1$.
%(see Appendix(\ref{app:Complexfermionlanguage})) and choose the inter-chain couplings: $S_{13},A_{14}$ along with the field parameter $h$ as variable parameters. 
Thus, we have a 3-dimensional parameter space of $\{h,\,A_{14},\,S_{13}\}$ to study the various phases and phase transitions. We also compute the winding number for the phases for the topological nature. The phases with non-zero winding number are topologically non-trivial and  the critical lines which are accompanied by a change of $W$ are then the topological phase transitions between the different topological phases.
Also, to compare the model to Fermionic chains, one can transform the Hamiltonian to complex Fermion basis (see Appendix \ref{app:Complexfermionlanguage}). In this form, it is can be seen that with the reflection symmetry imposed (eqn.(\ref{reflection})), the parameter $A_{ij}$ is like the hopping parameter and $S_{ij}$ behaves as the superconducting pairing. \newline
\hspace*{2mm} The first parameter plane we analyze is the $S_{13}-h$ plane  with topological classification with two qualitatively different fixed values of  interchain hopping $A_{14}=4,8$ (\Cref{fig:a14=4,fig:a14=8}).
For the case of  lower value of  $A_{14}=4$ (Fig.(\ref{fig:a14=4})), we observe 4 different phases with $W=\,-2,\,\pm1,\,0$. There are three critical curves, blue and cyan corresponding to $k=0$ transitions and green for incommensurate k, and 3 multicritical points labeled by A, B and C.
For the case of larger  $A_{14}=8$ (Fig.(\ref{fig:a14=8}) we additionally observe a phase with W= 2 (White) taking the total number of distinct topological phases to 5 and a $k=\pm\pi$ gap closing (Red critical line).% at $h=-h_{c1}$.\begin{center}
  %R \begin{center}
    \newline
\hspace*{2mm}Similarly we have also obtained the the phase diagrams in the $A_{14}-h$ parameter plane with fixing the interchain superconducting pairing coupling,  $S_{13}=4,8$  (see Fig.(\Cref{fig:s13=4,fig:s13=8})). For $S_{13}=4$ (Fig.(\ref{fig:s13=4})) we see phases with $W=-2,-1,0$ and two multicritical points labelled as A and B and for $S_{13}=8$ (Fig.(\ref{fig:s13=8})) we see phases with $W=\pm2,\pm1,0$ and 4 multicritical points labelled as A, B, C and D as Fig.(\ref{fig:a14merged}). The commensurate k critical lines seen for fixed $A_{14}$ cases are now replaced by curves in the fixed $S_{13}$ regime% as $A_{14}$ is a variable parameter 
(eqn.(\ref{eqn:criticalh})).%\begin{center}
    
\begin{figure}
    \begin{subfigure}{0.4\textwidth}
    \includegraphics[width=\textwidth,scale=0.3]{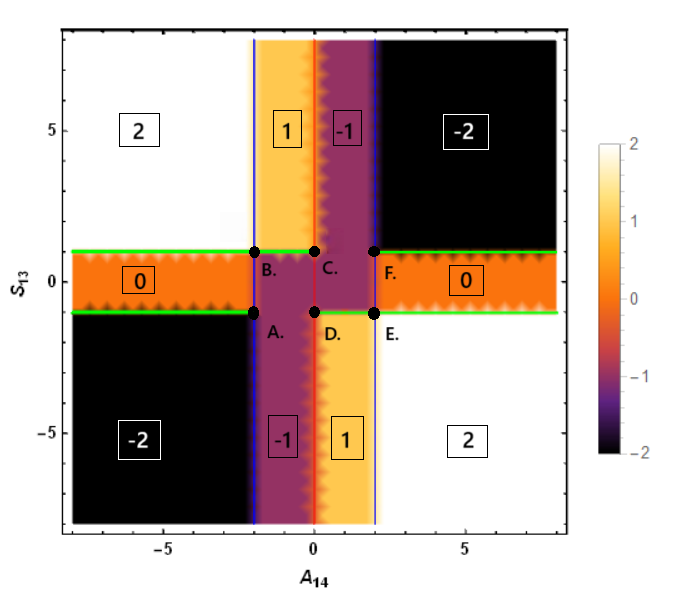}
    \caption{$h$=1.}

    \label{subfig:h=1}
\end{subfigure}
\begin{subfigure}{0.4\textwidth}
    \includegraphics[width=\textwidth,scale=0.3]{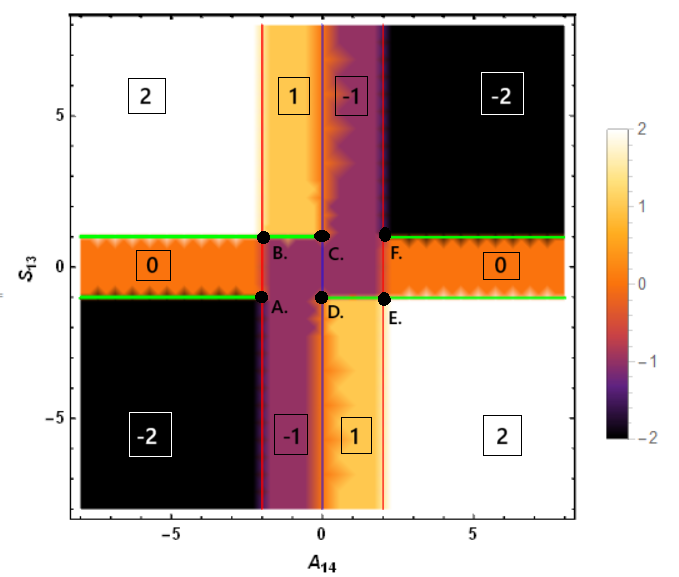}
    \caption{$h$=-1.} %Curves: Red: -$h_{c1}$ (k=$\pm\pi$), Blue: $h_{c2}$(k=0), Neon Green: incommensurate k. MCPs: A, B,C}
    \label{subfig:h=-1}
\end{subfigure}
  %\captionsetup{width=0.47\textwidth} % Set the total caption width
    \caption{\justifying
        Phase diagram in the \( A_{14} \)–\( S_{13} \) plane, with a) \( h = 1 \) and b) \( h = -1 \) and all other couplings set to unity. 
        The colors and numbers inside the boxes represent \( W \). 
        Blue (\( k = 0 \)), Red (\( k = \pm\pi \)), and Green (incommensurate \( k \)) are the critical curves. 
        The Multi critical points are labeled with the letters A, B, C, D, E and F.
    }

%  \caption{\begin{minipage}{0.4\textwidth}\raggedleft Density plot for \( W \) in the \( A_{14} \)-\ \( S_{13} \) plane, with  a)h=1 and b)h=-1. The colors and numbers inside the boxes represent \( W \). Blue (\(k=0\)), Red(\(k=\pm\)), and Green (incommensurate \( k \)) are the critical curves. The MCPs are labeled with the letters A, B, C, D, E and F.\end{minipage}}
        \label{fig:a14-s13h=1,-1}
    \end{figure}

%\end{center}%\newline
\hspace*{2mm}The last slicing possible are the $A_{14}-S_{13}$ parameter planes with $h$ as a fixed parameter and all other parameters set to unity(Fig.\ref{fig:a14-s13h=1,-1}).We find similar results to those previously reported. We observe that upon $h\to-h$ (\Cref{subfig:h=1,subfig:h=-1}), the topological order of the phases remains invariant. The values of the parameters are chosen to access qualitatively different phases with the exact choice of the values having no particular significance. %(see Fig.\ref{fig:hminushdensity}) consolidating the statement made in Sec.(\ref{subsec:Discrete symmtries}).
\newline\hspace*{2mm} As expected from the symmetry based topological classification  (Sec. \ref{subsec:Discrete symmtries}), across all the parameter planes, we find the  winding number  $W\in\mathds{Z}$. An interesting thing to note is that since the original model is exactly solvable for any value of $d$, we can expect to have larger values of $W$ in the same model. 
%In contrast to the single Kitaev chain model with $W\in\mathds{Z}_2$(see Ref.\cite{AYuKitaev_2001}) and models with next to nearest neighboring interactions (Ref.\cite{PhysRevB.85.035110}) or interacting chain models (pseudo 1 dimensional)(Refs.\cite{PhysRevB.89.174514,WU20123530,ZHOU20172426})) $W\in\mathds{Z}$ , we observe a larger range of allowed values of $W$ for a 1D model with only the nearest neighboring interactions. This novelty can be credited to the larger degrees of internal freedom on each site, consequently offering us a larger competing parameter space.%\begin{center}
%\end{center}
\subsection{\label{subsec:Spatial Distribution of MZE}Spatial Distribution of MZE}
\begin{figure}[!h]
\includegraphics[scale=0.35]{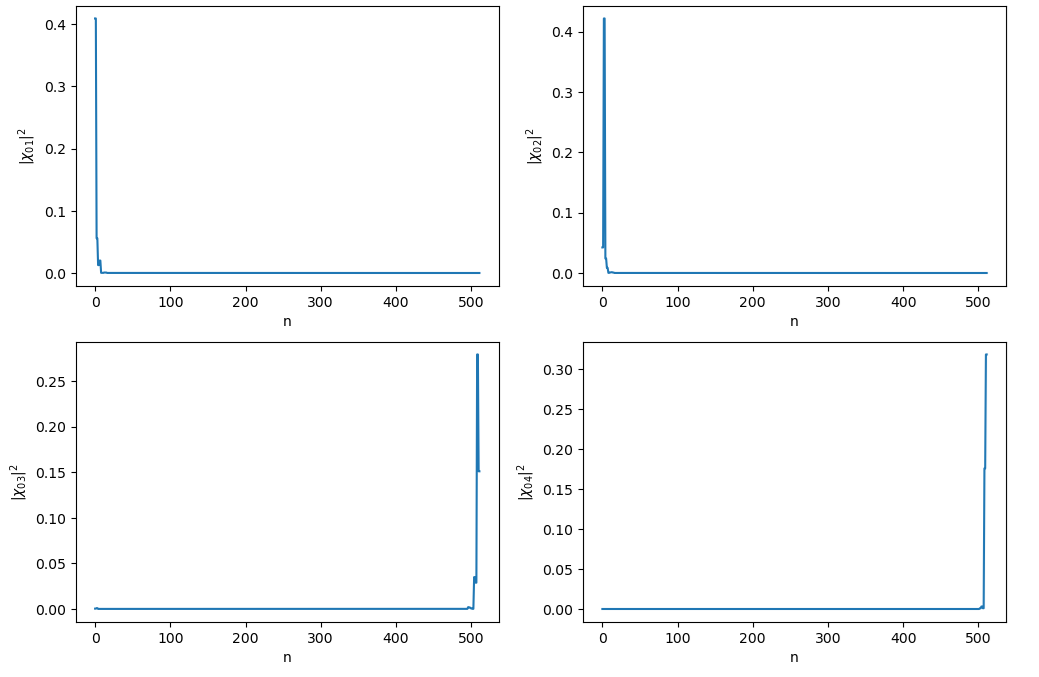}
       
        \caption{\justifying 
$|\chi_{01}|^2$, $|\chi_{02}|^2$, $|\chi_{03}|^2$, $|\chi_{04}|^2$ vs n for $A_{14}=8$ with $(S_{13},h)=(7,-1)$ corresponding to the $W=-2$ phase in the $S_{13}-h$ parameter plane (Fig.(\ref{fig:a14=8})). All the four MZEs ($\chi_{0i}$'s) decay exponentially into the bulk.}            \label{fig:expodecay}
        \end{figure}

\begin{figure}[!h]
\includegraphics[scale=0.4]{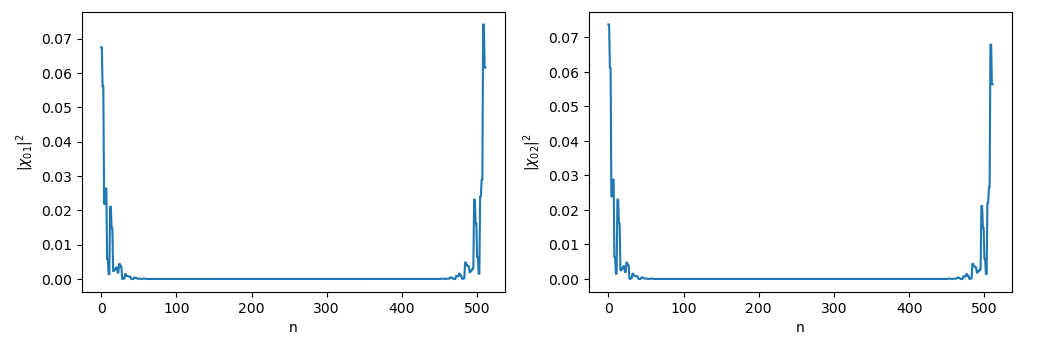}
       
        \caption{\justifying $|\chi_{01}|^2$, $|\chi_{02}|^2$ vs n for $S_{13}=8$ with $(A_{14},h)=(-8,-5)$ corresponding to the $W=1$ phase in the $A_{14}-h$ parameter plane (Fig.(\ref{fig:s13=8})). Both the MZEs ($\chi_{0i}$'s) display oscillatory exponential decay into the bulk.
        }
            \label{fig:osciexpodecay}
        \end{figure}

\begin{figure}[!h]
\includegraphics[scale=0.2]{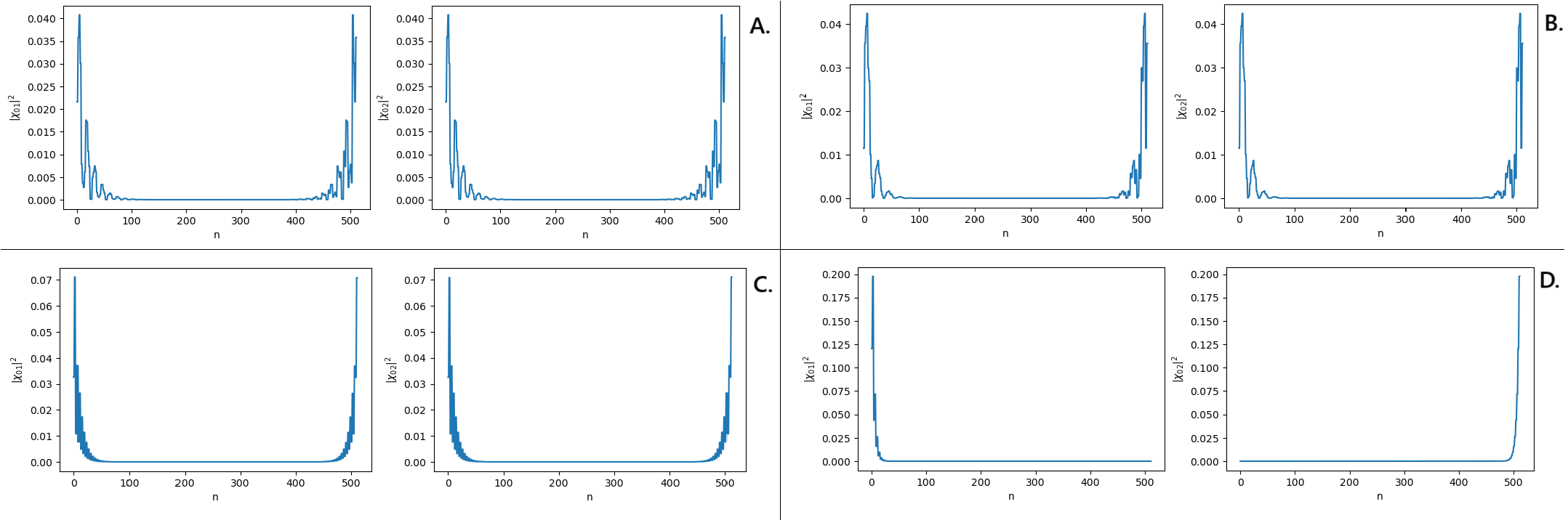}
       
        \caption{\justifying $|\chi_{01}|^2$, $|\chi_{02}|^2$ vs n for $S_{13}=4$,  $h=-5.5 $ with A.) $A_{14}= -4,\, B.)\,A_{14}=-2,\,C.)\,A_{14}=2,\,D.)\,A_{14}=4$ corresponding to the $W=-1$ phase 
        in the $A_{14}-h$ parameter plane (Fig.(\ref{fig:s13=4})). For all $A_{14}<0$, Both MZEs ($\chi_{0i}$'s) display and oscillatory exponential decay into the bulk but for increasing positive values of $A_{14}$ we begin to observe an exponentially decaying modes.
        }
            \label{fig:exception}
        \end{figure}

For topological phases with non zero winding number, we expect boundary localized zero energy states in an open chain of the model. For obtaining the nature of the bound states, we numerically solve for the zero energy eigen states of the  Majorana Hamiltonian in real space with open boundary conditions. 
We find that the number of zero energy modes of the system in different phases is equal to $2*|W|$. We label these zero energy modes as "$\chi_{0i}$" and since Max$(|W|)=2$, $i\in[1,4]$ and we have confirmed all these modes to be mutually orthogonal. We also confirm that these modes are localised at the edges by analyzing the spatial distribution of the probability density of the modes ($|\chi_{0i}|^2$). Most of these modes decay exponentially into the bulk, but a few display an oscillatory nature on top of this exponential decay.\newline
\hspace*{2mm} We consider a lattice with $N=128$ sites. Since each site hosts 4 Majorana fermions, the Hamiltonian matrix is of ($512 \times 512$) size. To label the $512$ Majorana modes, we use an index $n\in[0,511]$ and analyze the plots of $|\chi_{0i}|^2\,\,vs\,\,n$. In this section, we continue our analysis with $C_1=4$ and $C_2=6$.
Well within the phases, the Majorana edges are well localized and decay in the bulk. We find two types of decay behaviors, exponential decay (ED) and an oscillatory exponential decay (OED) as mentioned in the phase diagrams in Figs. (\ref{fig:a14merged}),(\ref{fig:s13merged}). For the phase diagram in $A_{14}-S_{13}$ plane, the intra-chain parameters are set at critical values and the Majorana behaviour is mixed in some regions, so we are not showing it on the phase diagram. We plot some examples of the Majorana modes in Figs. (\ref{fig:expodecay}),(\ref{fig:osciexpodecay}) and (\ref{fig:exception}). Similar behavior has been seen in other models, \textit{e.g.} Ref. \cite{PhysRevB.85.035110}, where it is seen that the oscillatory nature of the decay of the boundary modes is related to oscillatory nature of equal time two point correlations in the system. We haven't yet studied this connection for our model. It will be interesting to understand this connection in detail in future work. 

        %%%%%%

%%%%%%%%%%%%%%%%%%%%%%%%%%%%%%%%%%%%%%%%%%%%%%%%%%%%%%%%%%%%%%%%
\section{\label{sec:Critical Behaviour}Critical Behaviour}
In this section we explore the critical behaviour of the system in more detail. At the critical point, the excitation gap vanishes. The nature of this gap closing can be used to extract two of the critical exponents ($\nu$ and dynamical exponent $z$). These exponents are related to the correlation length $\xi$ via $\xi\sim |p-p_c|^{-\nu}$ and correlation time  $\xi_\tau \sim \xi^z$.  The momentum dependence of the excitation energy $\epsilon_-(k)$ near the gap closing point $k_c$ is given by $\epsilon_-\sim |k-k_c|^z$. Also as we approach the critical point by varying any of the parameters of the system (denoted by $p$), $\epsilon_-\sim |p-p_c|^{\nu z}$ \cite{Sachdev2011-gv,Dutta2015-ke}. Analyzing the behavior of $\epsilon_-$ near criticality, we characterize the critical lines and the multi-critical points of the model. 
We encounter two distinct kinds of scaling, possibly hinting at two distinct universality classes.  

%%%%%%%%%%%%%%%%%%%%%%%%%
\subsubsection{Ising Universality Class}
\begin{figure}
% \begin{center} 
%\begin{figure}[h]
\begin{overpic}[width=0.4\textwidth,,height=3.75cm]{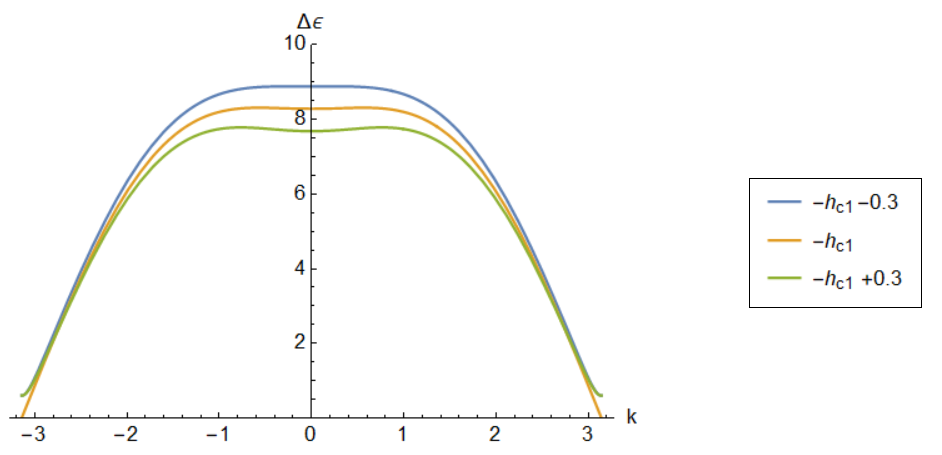}\put(38,22){
\frame{\includegraphics[width=0.15\textwidth]{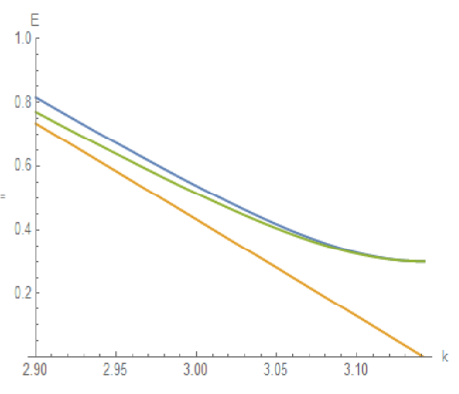}}}%
\put(15,45){(a)}%
\put(65,45){(b)}%
\end{overpic}
\captionof{figure}{\justifying a) $\Delta\epsilon_{-}$ vs k for $S_{13}=8$, $A_{14c}=7$ at $\{-h_{c1},\,-h_{c1}\mp0.3\}$ b)Inset: Zoomed in around $k=\pi$}
    \label{fig:kpi S13=8} % Label for the main figure
%\end{center}
\end{figure}

\begin{figure}
% \begin{center} 
%\begin{figure}[h]
\begin{overpic}[width=0.4\textwidth,height=3.75cm]{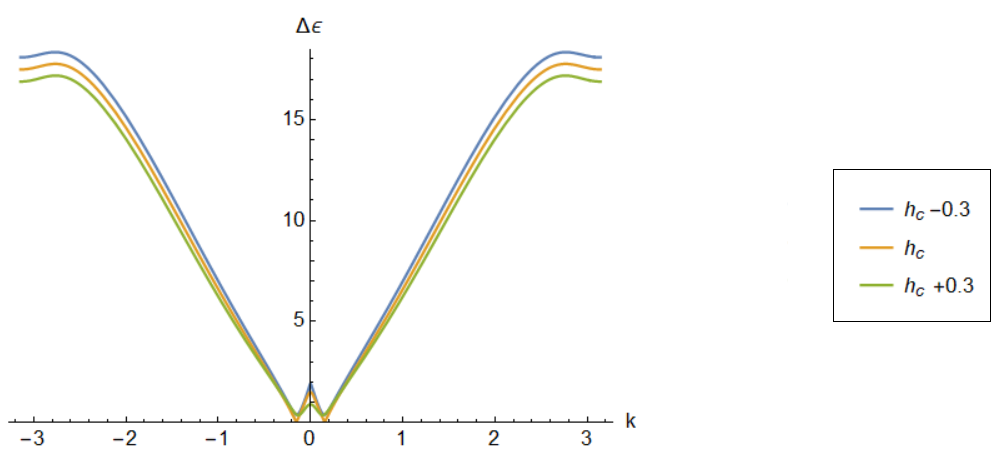}\put(42,22){
\frame{\includegraphics[width=0.15\textwidth]{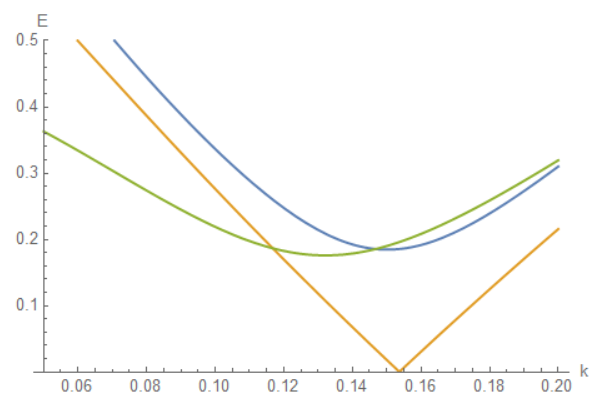}}}%
\put(15,40){(a)}%
\put(70,40){(b)}%
\end{overpic}
\captionof{figure}{\justifying a) $\Delta\epsilon_{-}$ vs k  for $S_{13}=8$,$\,\,A_{14c}=0$ at $\{h_{c}=-3\sqrt{5/2},\,h_{c}\mp0.3\}$  b)Inset: Zoomed in around $k_c$}
\label{fig:incommesurate k s13=8} % Label for the main figure
%\end{center}
\end{figure}

\begin{figure}
 %\begin{center} 
%\begin{figure}[h]
\begin{overpic}[width=0.5\textwidth,,height=3.75cm]{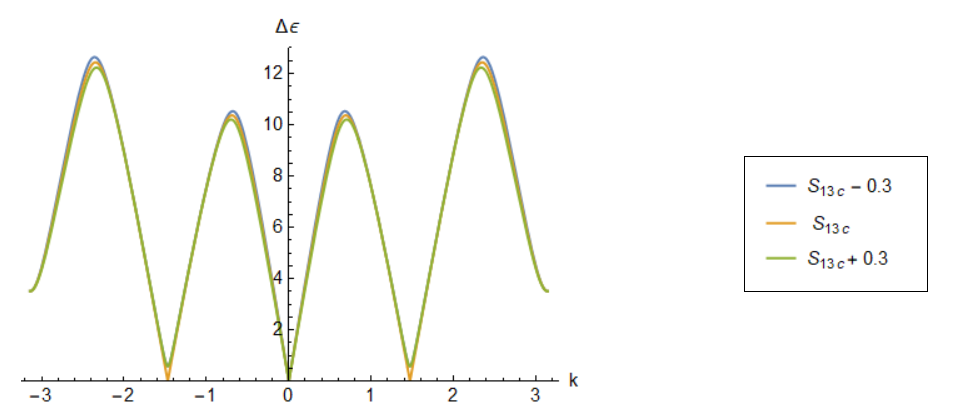}\put(38,19){
\frame{\includegraphics[width=0.18\textwidth]{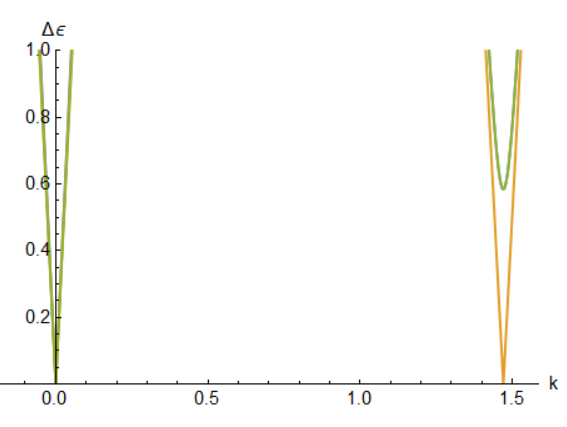}}}%
\put(15,40){(a)}%
\put(55,40){(b)}%
\end{overpic}
\captionof{figure}{\justifying a) $\Delta\epsilon_{-}$ vs k, around MCP A for $A_{14}=4$, at $\{S_{13c},S_{13c}\pm0.3\}$. b)Inset: Zoomed in around $k=0,k_c$}
    \label{fig:MCP A a14=4} % Label for the main figure
%\end{center}
\end{figure}

\hspace*{2mm} %In Eqn.(\ref{cricexp}), 
If $z=1$ and $\nu=1$, the critical point (CP) possibly belongs to the Ising Universality class (Refs. \cite{Dutta2015-ke,PhysRevB.85.035110,AYuKitaev_2001}. We found that all the Critical points/multi critical points with linear gap closings around $k_c$ belong to the above-mentioned universality class. Below, we discuss cases of such phase transitions. %The critical a) commensurate k gap closing ($k=0,\,\pm\pi$), b) Incommensurate k gap closing, c) all MCPs with linear gap closings around the respective $k_c$.\newline
%%%%%%%%%%%%%%
\paragraph{Commensurate k:\newline}
We consider a $k=\pm\pi$ gap closing occurring at the critical point ($A_{14c},-h_{c1}$)=$(7,-5 (-1+\sqrt{2})$ for $S_{13}=8$ (Cyan line in Fig(\ref{fig:s13=8}), separating the W=-2 and W=-1 phases. The energy gap  closes linearly around $k=\pm\pi$ at the critical point (Fig.\ref{fig:kpi S13=8}).
    \begin{equation}
    \epsilon_{-}(k)\sim|k\pm\pi|^1 \,\,\;\,\, \epsilon_{-}(p)\sim|p-p_c|^{1}
\end{equation}
Here, $p\in\{A_{14},h\}$. Thus, $z=1$ and $\nu=1$, pointing to the Ising Universality Class. All the commensurate k phase transitions of $A_{14}=4,8$ and $S_{13}=4,8$ display similar results.
\newline
%%%%%%%%%%%%%%%%%%%%%%
\paragraph{Incommensurate k:\newline}
We encounter an incommensurate k  gap closing at $k_c = \pm 0.153697$ at the critical point  $(A_{14c},h_{c})=(0,-3\sqrt{5/2})$ for $S_{13}=8$ (Green line in Fig.(\ref{fig:s13=8})), separating the W=-1 and W=1 phases. Similar to the commensurate cases, we obtain $\nu=1$, $z=1$ %, for the linear gap closing around $k_c$'s 
(Fig.(\ref{fig:incommesurate k s13=8})).  All the incommensurate k phase transitions of $A_{14}=4,8$ and $S_{13}=8$ share the same result.\newline 
%%%%%%%%%%%%%%%%%%%%
\paragraph{Multi-critical point  A of $A_{14}=4$ phase diagram (Fig. \ref{fig:a14merged}a):\newline} %We have analayzed  %with the coordinates
%$(S_{13c},h_{c1})=(−4.9022,-5+\sqrt{17})$ 
%(see Fig.(\ref{fig:a14=4}))%. This MCP is 
%formed at the intersection of $k=0$ and an incommensurate k curve. 
As plotted in Fig. (\ref{fig:MCP A a14=4}), the energy band gap of this multi-critical point closes linearly around the critical values  $k_c=0,\pm 1.47009$. Similar to the previous cases, we obtain $z=1$ and $\nu=1$.
%We obtain the same set of CP as before, $z=1$ and $\nu=1$. 
multi-critical points A and B ($k_c=0,\pm\pi$) of $S_{13}=4,8$ (\Cref{fig:s13=4,fig:s13=8}) and multi-critical point A of $A_{14}=8$ (Fig.(\ref{fig:a14=8})) also show the same behavior.
%%%%%%%%%%%%%%%%%%%%%%%%
\subsubsection{Lifshitz Points}
\begin{figure}
 %\begin{center} 
%\begin{figure}[h]
\begin{overpic}[width=0.5\textwidth,,height=3.75cm]{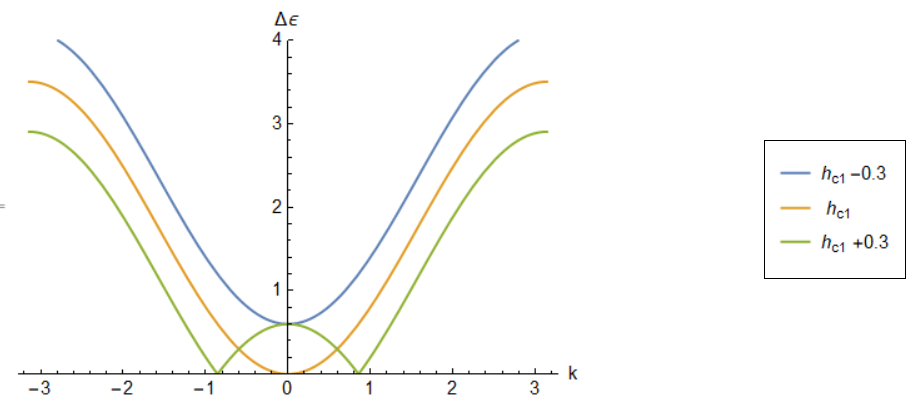}\put(45,20){
\frame{\includegraphics[width=0.18\textwidth]{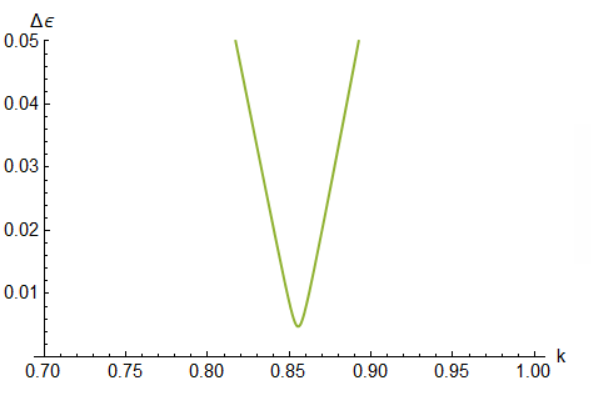}}}%{a14=4,MCP C, zoomed}}}%
\put(15,40){(a)}%
\put(75,40){(b)}%
\end{overpic}
\captionof{figure}{\justifying a) $\Delta\epsilon_{-}$ vs k, around MCP B for $A_{14}=4$, at $\{h_{c1},h_{c1}\mp0.3\}$. b)Inset: Zoomed in plot shows $\Delta\epsilon_{-}$ stays open around this MCP.}
    \label{fig:MCP B A14=4} % Label for the main figure
%\end{center}
\end{figure}

The model also displays another distinct scaling behavior at a few of the multi-critical points. As shown in Fig. \ref{fig:MCP B A14=4}, the energy band gap closes quadratically (Orange) around multi-critical point B of $A_{14}=4$ (Fig.(\ref{fig:a14=4})). %It is formed at $(S_{13c},h_{c2})=[-\frac{1}{4}(1+5\sqrt{17}),-5-\sqrt{17}]$, intersection of an incommensurate k and a k=0 critical curve. 
 %indicates that the gap closes quadratically around $k=0$ at MCP C (Orange) 
% and stays open for values of $h$ around it. 
Similar quadratic gap closings are seen for all %Conclusively, as one moves towards an 
multi-critical points formed at the termination of the incommensurate k curve at a commensurate curve. As we approach such MCPs along an incommensurate critical curve, the two linear gap closings at $\pm k_c$ (Fig.(\ref{fig:incommesurate k s13=8})) move closer and eventually coalesce at $k=0$ or $k=\pm \pi$ (depending on the commensurate curve). For these points, %it terminates at.
\begin{equation}
    \epsilon_{-}(k)\sim|k|^2\,\,;\,\, \epsilon_{-}(p)\sim|p-p_c|^1
\end{equation}
here, $p\in\{S_{13},h\}$. Thus $z=2$ and $\nu=\frac{1}{2}$. These transition might be  Lifshitz points (Refs.\cite{Sugimoto_2015,JP,Huang_2022}).

%\textcolor{red}{\textbf{See Ref{\cite{Huang_2022}}}}?

%%%%%%%%%%%%%%%%%%%%%%%%%%%%%%%%%%%%%%%%%%
\subsection{\label{subsec:Non-analyticities of GS}Non-analyticy of GS Energy at the critical points}
\begin{figure}
    \begin{subfigure}{0.5\textwidth}
\includegraphics[width=\textwidth,scale=0.3,height=3.75cm]{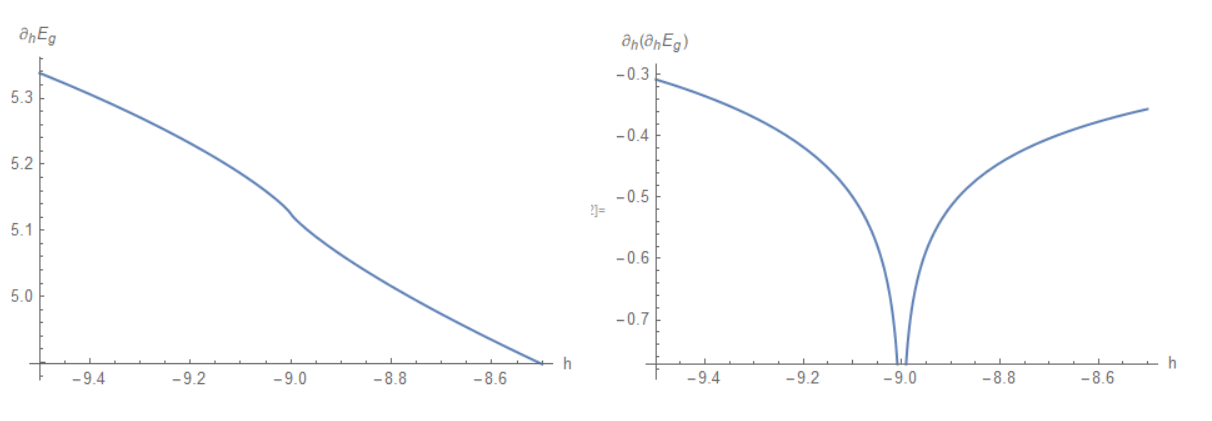}
    \label{subfig:Egdel1}
\end{subfigure}
\begin{subfigure}{0.5\textwidth}
    \includegraphics[width=\textwidth,scale=0.3,height=3.75cm]{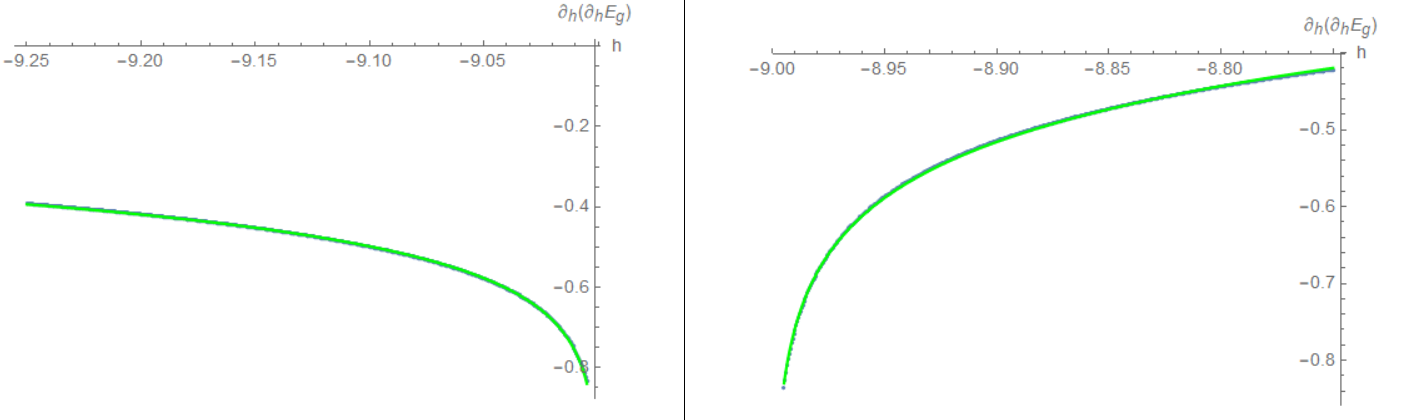}
    \label{subfig:Egdel1fit}
\end{subfigure}
      \caption{\justifying Top: the behavior of $\frac{\partial E_g}{\partial h}$ and $\frac{\partial^2E_g}{\partial h^2}$. Bottom: Fitting of $\frac{\partial^2E_g}{\partial h^2}$ (blue)  to the function $a*log(b|h-h_c|$ (green) for the critical point  $(A_{14c}=\sqrt{15},h_{c}=-9)$ with $S_{13}=4\,$. The fit parameters are Left: $a=0.114081,b =0.127172$ and Right: $a=0.104806,b=-0.072978$ }
        \label{fig:Eg1}
    \end{figure}

We can further look at the phase transitions via the non-analyticity of the the ground state energy. From equation \ref{GS1}, we can reformulate $E_{gs}$ in terms of $F_k\,\,\text{and}\,\,G_k$ (Ref.\cite{PhysRevE.106.024114,PhysRevB.80.014414ZVYAGIN}) as, 
\begin{equation}
E_{g}=-\sqrt{2}\sum_{k\geq 0} \sqrt{F_{k}+\sqrt{G_{k}}}\label{GS2}
\end{equation}
The analytic behaviour of $E_{g}$ and its derivatives around all distinct cases of topological phase transitions is that of a $2^{nd}$ order phase transitions. 
For all Ising type phase transitions \textit{i.e.} the critical curves with $\Delta W=\,1,\,2$ and the multi-critical points with $z=1,\,\nu=1$,  the $2^{\text{nd}}$ derivative of ground state energy  $\partial^2 E_{g}/\partial p^2\,\,;\,\,p\in\{h,S_{13},A_{14}\}$ display a logarithmic divergence ($ a\,log(b*|p-p_c|)$, with a and b being fit parameters)
%around the CP (Ref.\cite{LIU2021126122}). 
This is illustrated in fig.\ref{fig:Eg1} for the critical point $(A_{14c}=\sqrt{15},h_{c}=-9 $ with $S_{13}=4$ corresponding to a phase transition with $\Delta W=1$ (Fig.\ref{fig:s13=4}).  %corresponding to a phase transition with . 
    
    \begin{figure}
    \begin{subfigure}{0.5\textwidth}
\includegraphics[width=\textwidth,scale=0.3,height=3.75cm]{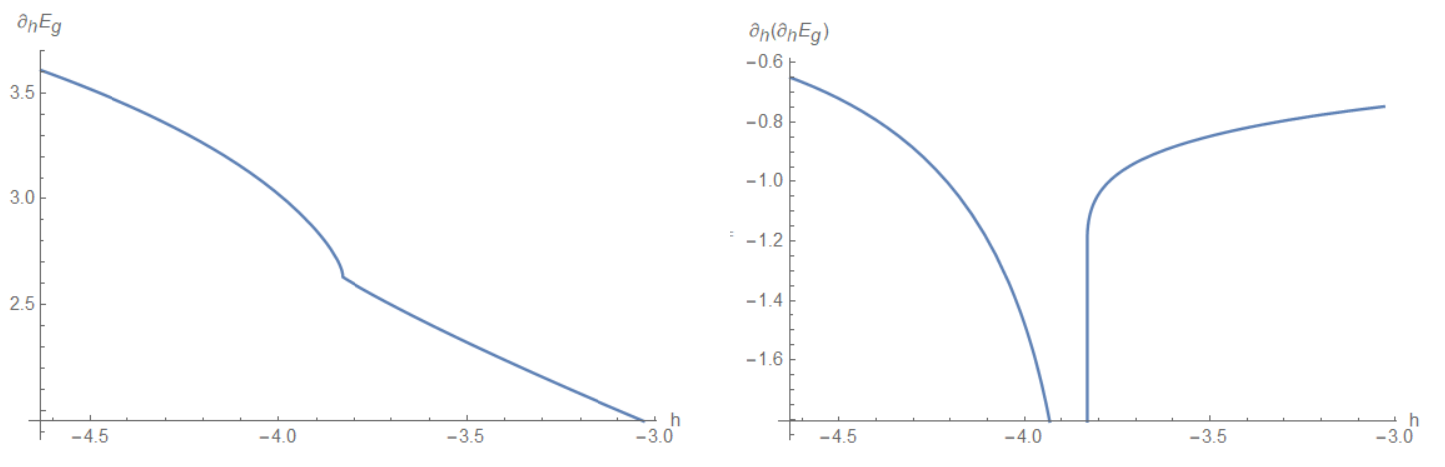}

    \label{subfig:Egmcpc}
\end{subfigure}
\begin{subfigure}{0.5\textwidth}
    \includegraphics[width=\textwidth,scale=0.3,height=3.75cm]{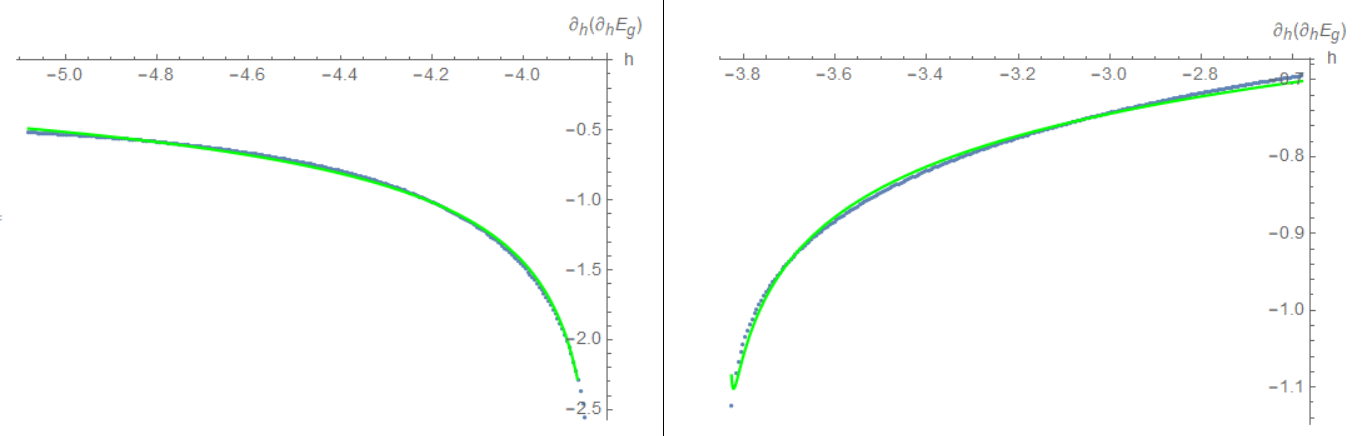}
    \label{subfig:Egmcpcfit}
\end{subfigure}
      \caption{\justifying Top: the behavior of $\frac{\partial E_g}{\partial h}$ and $\frac{\partial^2E_g}{\partial h^2}$. Bottom: Fitting of $\frac{\partial^2E_g}{\partial h^2}$ (blue)  to the function $a*\log(b/|h-h_c|)*(1/|h-h_c|^c)$ (green) for the critical point  $(A_{14c}=\frac{2}{39} \left(5 \sqrt{10}-4\right),h_c=-3.83)$ with $S_{13}=8$.The fit parameters are Left: $a=-0.306114,b =6.51208,c=0.145741$ and Right: $a=0.0852256,b=2.02778\,\times\,10^{-4},c=0.260431$ }
        \label{fig:Eg2}
    \end{figure}
    
However, for the multi-critical points with $z=2,\,\nu=1$, the behavior is captured by $\partial^2 E_{g}/\partial p^2\sim a\,log(b*|p-p_c|)*1/|p-p_c|^c,\,\,$ with a, b and c being the fit parameters. This is illustrated in Fig.(\ref{fig:Eg2})for the MCP C, of the $S_{13}=8$ phase diagram (Fig. \ref{fig:s13=8}).

%%%%%%%%%%%%%%%%%%%%%%%%%%%%%%%%%%%%%%%%%%%%%%%%%%%%%%%%%%%%%%%%%%
%\section{\label{sec:Correlations}
%Correlations}

%%%%%%%%%%%%%%%%%%%%%%%%%%%%%%%%%%%%%%%%%%%%%%%%%%%%%%%%%%%%%%%%%
\section{\label{sec:Conclusions}Conclusions}
The $\Gamma-$matrix one dimensional model displays a rich phase diagram with a large parameter space to tune. For the $4$ dimensional matrix model, we found topological phases with winding number allowed to take values $|W|=0,1,2$ belonging to the C-II class of the symmetry protected topological phases. Since the model is exactly solvable for larger values of $\Gamma$ matrix dimensions also, we could find lager number of topological phases in a local nearest neighbour interactions model. In agreement with the topological classification, we also find topologically protected localized Majorana modes on the boundaries of open chains. We have also characterized the quantum critical points with scaling behaviour and find interesting behaviour at some of the multi-critical points. 
In the picture of the model as two interacting spin chains, it could be related to spin ladder systems and could capture properties of spin liquid phases in such system. We leave the study of the nature of correlation functions and comparison to spin ladder models for future work. The model is also expected to have large quantum critical regions due to extended quantum critical fans along the critical lines \cite{PhysRevB.108.155143} which would also be an interesting direction to explore.

\begin{acknowledgments}
We thank Prithvi Narayan P  for collaboration during initial phases of the work and fruitful discussions throughout. We also thank Subhro Battacharjee  and Sumathi Rao for helpful insights.
\end{acknowledgments}

\appendix

\section{Complex Fermion Language\label{app:Complexfermionlanguage}}
We define complex fermion creation and annihilation operators in k-space as follows (Ref.(\cite{PhysRevE.106.024114})),
\begin{equation}
    c^i_{k}=\frac{\chi^{2i-1}_k+i\chi^{2i}_k}{\sqrt{2}}\,\,;\,\,c^{i\dag}_{-k}=\frac{\chi^{2i-1}_k-i\chi^{2i}_k}{\sqrt{2}}\,\,\label{Majorana to complex trans}
\end{equation}
The form of the Hamiltonian $h_{CF}(k)$ in the basis $C_k=\begin{pmatrix}
        c^{1\dag}_{-k}&c^{2\dag}_{-k}&c^{1}_{k}&c^{2}_{k}
    \end{pmatrix}^T$ is ,\scriptsize
\begin{equation}
\begin{pmatrix}
 -h_1-\cos (k) A_{12} & -\cos (k)A_{14} & \sin (k) S_{11} & \sin (k) S_{13} \\
 -\cos (k)A_{14} & -h_2-\cos (k) A_{34} & \sin (k)S_{13} & \sin (k) S_{33} \\
 \sin (k) S_{11} & \sin (k)S_{13} & h_1+\cos (k) A_{12} & \cos (k)A_{14} \\
 \sin (k)S_{13} & \sin (k) S_{33} & \cos (k)A_{14}& h_2+\cos (k) A_{34} \\
\end{pmatrix}\label{complexhamiltonian1111}    
\end{equation}
\normalsize
The Unitary transformation matrix that facilitates this is,
    \begin{equation}
       U_t=\frac{1}{\sqrt{2}} \left(
\begin{array}{cccc}
 1 & -i & 0 & 0 \\
 0 & 0 & 1 & -i \\
 1 & i & 0 & 0 \\
 0 & 0 & 1 & i \\
\end{array}
\right)\label{utcm1}
    \end{equation}
\normalsize
If one choose the basis $\tilde{C}_k=(c^{1\dag}_{-k},c^{1}_{k},c^{2\dag}_{-k},c^{2}_{k})^T$ instead of $C_k$, the form of $h_{CF}(k)$ resemebles that of two interacting Kitaev chains (See Ref.\cite{AYuKitaev_2001,PhysRevB.89.174514,WU20123530,ZHOU20172426}). In this case, the upper and the lower $2\times2$ diagonal blocks represent the individual chain Hamiltonians and the off-diagonal blocks represent the interaction between the two chains. In this basis we can identify the $A_{\mu\nu}$'s to be like hopping parameters and $S_{\mu\nu}$'s to be like superconducting pairing. %Here, $(A_{12},S_{11})$ and $(A_{34},S_{33})$ belong to $\text{chain}_1$ and $\text{chain}_2$ respectively. 
$A_{14},S_{13}$  parameterize the inter-chain coupling terms.
\begin{comment}
\section{Commutation Relations:}
For d=2, the commutation relations of $\Gamma^{\mu}$'s with $\,\Gamma^{2d+1}$ are as followsa,
\begin{equation*}[\Gamma^1_a,\Gamma^5_a]=-2i(\sigma^2_1\otimes\mathrm{I}_2)_a\,\,;\,\,[\Gamma^2_a,\Gamma^5_a]=2i-2i(\sigma^1_1\otimes\mathrm{I}_2)_a
\end{equation*}
\begin{equation*}
    [\Gamma^4_a,\Gamma^5_a]=2i(\mathrm{I}_1\otimes\sigma^1_2)_a \,\,;\,\,[\Gamma^3_a,\Gamma^5_a]=-2i(\mathrm{I}_1\otimes\sigma^2_2)_a\end{equation*}
Similarly, the commutation relations between the $\Gamma^{\mu}$'s among themselves,
\begin{equation*}
[\Gamma^1_a,\Gamma^2_a]=2i(\sigma^3_1\otimes\mathrm{I}_2)_a\,\, ;\,\, [\Gamma^3_a,\Gamma^4_a]=2i(\mathrm{I}_1\otimes\sigma^3_2)_a 
\end{equation*}
\begin{equation*}
[\Gamma^1_a,\Gamma^4_a]=-2i(\sigma^2_1\otimes\sigma^2_2)_a\,\, ;\,\, [\Gamma^2_a,\Gamma^3_a]=2i(\sigma^1_1\otimes\sigma^1_2)_a 
\end{equation*}
\begin{equation*}
[\Gamma^1_a,\Gamma^3_a]=-2i(\sigma^2_1\otimes\sigma^1_2)_a\,\, ;\,\, [\Gamma^2_a,\Gamma^4_a]=2i(\sigma^1_1 \otimes \sigma^2_2)_a
\end{equation*}

\end{comment}

\bibliography{ref1}
%\bibliography{apssamp}
% Produces the bibliography via BibTeX.

\end{document}